\documentclass[prl,article,nofootinbib,twocolumn,preprintnumbers,superscriptaddress]{revtex4}
\pdfoutput=1

\usepackage{amsmath}
\usepackage{graphicx}

\usepackage[utf8]{inputenc}

\usepackage[usenames,dvipsnames]{color}
\definecolor{MyGrey}{rgb}{0,0,0} 
\definecolor{MyDarkBlue}{rgb}{0.,0.,1} 
\definecolor{MyLightBlue}{rgb}{0.22,0.51,0.9}
\usepackage[colorlinks=true,linkcolor=MyDarkBlue,citecolor=MyDarkBlue,urlcolor=MyLightBlue,bookmarksnumbered=true,bookmarksopen]{hyperref}

\newcommand{\eps}{\ensuremath{\epsilon}}
\newcommand{\Dmq}{\ensuremath{\Delta m^2}}
\newcommand{\mysect}[1]{\par\smallskip\textit{#1}}

\begin{document}

\preprint{FERMILAB-PUB-17-308-T, YITP-SB-17-28, IFT-UAM/CSIC-17-073} 

\title{A COHERENT enlightenment of the neutrino Dark Side }

\author{Pilar Coloma}
\email{pcoloma@fnal.gov}
\affiliation{Theory Department, Fermi National Accelerator Laboratory,
  P.O.\ Box 500, Batavia, IL 60510, USA}

\author{M.~C.~Gonzalez-Garcia,}
\email{maria.gonzalez-garcia@stonybrook.edu}
\affiliation{Departament de Fis\'{\i}ca Qu\`antica i Astrof\'{\i}sica
  and Institut de Ciencies del Cosmos, Universitat de Barcelona,
  Diagonal 647, E-08028 Barcelona, Spain}
\affiliation{Instituci\'o Catalana de Recerca i Estudis Avan\c{c}ats
  (ICREA), Pg.\ Lluis Companys 23, 08010 Barcelona, Spain.}
\affiliation{C.N.~Yang Institute for Theoretical Physics, Stony Brook
  University , Stony Brook, NY 11794-3840, USA}

\author{Michele Maltoni,}
\email{michele.maltoni@csic.es}
\affiliation{Instituto de F\'{\i}sica Te\'orica UAM/CSIC, Calle de
  Nicol\'as Cabrera 13--15, Universidad Aut\'onoma de Madrid,
  Cantoblanco, E-28049 Madrid, Spain}

\author{Thomas Schwetz}
\email{schwetz@kit.edu}
\affiliation{Institut f\"ur Kernphysik, Karlsruher Institut f\"ur
  Technologie (KIT), D-76021 Karlsruhe, Germany}

\begin{abstract} 
  In the presence of non-standard neutrino interactions (NSI),
  oscillation data are affected by a degeneracy which allows the solar
  mixing angle to be in the second octant (\emph{aka} the dark side)
  and implies a sign flip of the atmospheric mass-squared
  difference. This leads to an ambiguity in the determination of the
  ordering of neutrino masses, one of the main goals of the current
  and future experimental neutrino program. We show that the recent
  observation of coherent neutrino--nucleus scattering by the COHERENT
  experiment, in combination with global oscillation data, excludes
  the NSI degeneracy at the $3.1\sigma$ ($3.6\sigma$) CL for NSI with
  up (down) quarks.
\end{abstract}


\maketitle

The standard three-flavour oscillation scenario is supported by a
large amount of data from neutrino oscillation experiments. The
determination of oscillation parameters (see, \textit{e.g.},
Ref.~\cite{Esteban:2016qun}) is very robust, and for a broad range of
new physics scenarios only small perturbations of the standard
oscillation picture are allowed by data. There is, however, an
exception to this statement: in the presence of non-standard neutrino
interactions (NSI)~\cite{Wolfenstein:1977ue, Valle:1987gv,
  Guzzo:1991hi} a degeneracy exists in oscillation data, leading to a
qualitative change of the lepton mixing pattern. This was first
observed in the context of solar neutrinos, where for suitable NSI the
data can be explained by a mixing angle $\theta_{12}$ in the second
octant, the so-called LMA-Dark (LMA-D)~\cite{Miranda:2004nb}
solution. This is in sharp contrast to the established standard MSW
solution~\cite{Wolfenstein:1977ue, Mikheev:1986gs}, which requires a
mixing angle $\theta_{12}$ in the first octant.

The origin of the LMA-D solution is a degeneracy in oscillation
probabilities due to a symmetry of the Hamiltonian describing neutrino
evolution in the presence of NSI~\cite{GonzalezGarcia:2011my,
  Gonzalez-Garcia:2013usa, Bakhti:2014pva, Coloma:2016gei}. This
degeneracy involves not only the octant of $\theta_{12}$ but also a
change in sign of the larger neutrino mass-squared difference,
$\Dmq_{31}$, which is used to parameterize the type of neutrino mass
ordering (normal versus inverted). Hence, the LMA-D degeneracy makes
it impossible to determine the neutrino mass ordering by oscillation
experiments \cite{Coloma:2016gei}, and therefore jeopardizes one of
the main goals of the upcoming neutrino oscillation program.  As
discussed in Refs.~\cite{Coloma:2016gei, Coloma:2017egw,
  Miranda:2004nb, Escrihuela:2009up}, non-oscillation data (such as
that from neutrino scattering experiments) is needed to break this
degeneracy.

Recently, coherent neutrino--nucleus scattering has been observed for
the first time by the COHERENT experiment~\cite{Akimov:2017ade}, using
neutrinos produced at the Spallation Neutron Source (SNS) in Oak Ridge
National Laboratory. The observed interaction rate is in good
agreement with the Standard Model (SM) prediction and can be used to
constrain NSI. In this Letter we show that this result excludes the
LMA-D solution at $3.1\sigma$ ($3.6\sigma$) CL for NSI with up (down)
quarks when combined with oscillation data.

\mysect{NSI formalism and the LMA-D degeneracy}.
We consider the presence of neutral-current (NC) NSI in the form of
dimension-six four-fermion operators, following the notation of
Ref.~\cite{Gonzalez-Garcia:2013usa}.  Since we are interested in the
contribution of the NSI to the effective potential of neutrinos in
matter, we will only consider vector interactions in the form
\begin{equation}
  \label{eq:L}
  \mathcal{L}_\text{NSI} = -2\sqrt{2} G_F \epsilon_{\alpha\beta}^{f,V}
  (\bar\nu_{\alpha L} \gamma^\mu \nu_{\beta L}) (\bar{f} \gamma_\mu f) \,,
\end{equation}
where, $\alpha, \beta = e,\mu,\tau$, and $f$ denotes a SM fermion. The
parameter $\epsilon_{\alpha\beta}^{f,V}$ parametrizes the strength of
the new interaction relative to the Fermi constant $G_F$, and
hermiticity requires that $\epsilon_{\alpha\beta}^{f,V} =
(\epsilon_{\beta\alpha}^{f,V})^*$.  In gauge invariant models of new
physics at high energies, NSI parameters are expected to be subject to
tight constraints from charged lepton observables~\cite{Gavela:2008ra,
  Antusch:2008tz}, leading to no visible effect in
oscillations. However, more recently it has been argued that viable
gauge models with light mediators (\textit{i.e.}, below the
electro-weak scale) may lead to observable effects in oscillations
without entering in conflict with other bounds~\cite{Farzan:2015doa,
  Farzan:2015hkd, Babu:2017olk} (see also Ref.~\cite{Miranda:2015dra}
for a discussion). In particular, for light mediators, bounds from
high-energy neutrino scattering experiments such as
CHARM~\cite{Dorenbosch:1986tb} and NuTeV~\cite{Zeller:2001hh} do not
apply. In this framework, prior to the COHERENT results, the only direct  
bounds on NC-NSI with quarks arise from their effect on neutrino
oscillations when propagating in matter (for bounds in the heavy
mediator case see~\cite{Coloma:2017egw}).  In the following we will
assume that the mediator responsible for the NSI has a mass larger
than about 10~MeV, and hence the contact interaction approximation
adopted in Eq.~\eqref{eq:L} applies for COHERENT.

The operators in Eq.~\eqref{eq:L} will contribute to the effective matter
potential in the Hamiltonian describing the evolution of the neutrino
flavour state:
\begin{align}
  \label{eq:Hmat}
  H_\text{mat}
  &= \sqrt{2} G_F N_e (x) 
  \begin{pmatrix}
    1 + \epsilon_{ee} & \epsilon_{e\mu} & \epsilon_{e\tau} \\
    \epsilon_{e\mu}^* & \epsilon_{\mu\mu} & \epsilon_{\mu\tau} \\
    \epsilon_{e\tau}^* & \epsilon_{\mu\tau}^* & \epsilon_{\tau\tau}
  \end{pmatrix},
  \\[3mm]
  \label{eq:eps-eff}
  \epsilon_{\alpha\beta}(x)
  &= \sum_{f=e,u,d} Y_f(x) \epsilon_{\alpha\beta}^{f,V} \,,
\end{align}
with $Y_f(x) \equiv N_f(x) / N_e(x)$, $N_f(x)$ being the density of
fermion $f$ along the neutrino path. Therefore, the effective NSI
parameters entering oscillations, $\epsilon_{\alpha\beta}$, may depend
on $x$ and will be generally different for neutrinos crossing the
Earth or the solar medium. The ``1'' in the $ee$ entry in
Eq.~\eqref{eq:Hmat} corresponds to the standard matter potential
\cite{Wolfenstein:1977ue, Mikheev:1986gs}.  In this work we consider
the cases of NSI with either up ($f=u$) or down ($f=d$) quarks.  Note
that oscillation experiments are only sensitive to differences between
the diagonal terms in the matter potential (for instance,
$\epsilon_{ee} - \epsilon_{\mu\mu}$ and $\epsilon_{\tau\tau} -
\epsilon_{\mu\mu}$).

In general, neutrino evolution is invariant if the relevant
Hamiltonian is transformed as $H \to -H^*$. This is a consequence of
the CPT symmetry, see~\cite{GonzalezGarcia:2011my,
  Gonzalez-Garcia:2013usa} for a discussion in the context of NSI.  In
vacuum this transformation can be realized by changing the oscillation
parameters as
\begin{equation}
  \label{eq:osc-deg}
  \begin{aligned}
    \Dmq_{31} &\to -\Dmq_{31} + \Dmq_{21} = -\Dmq_{32} \,, \\
    \sin\theta_{12} &\leftrightarrow \cos\theta_{12} \,,\\
    \delta &\to \pi - \delta \,,
  \end{aligned}
\end{equation}
where $\delta$ is the leptonic Dirac CP phase, and we are using here
the parameterization conventions from Ref.~\cite{Coloma:2016gei}.  The
symmetry is broken by the standard matter effect, which allows a
determination of the octant of $\theta_{12}$ and (in principle) of the
sign of $\Dmq_{31}$. However, in the presence of NSI, the symmetry can
be restored if in addition to the transformation
Eq.~\eqref{eq:osc-deg}, NSI parameters are transformed
as~\cite{Gonzalez-Garcia:2013usa, Bakhti:2014pva, Coloma:2016gei}
\begin{equation}
  \label{eq:NSI-deg}
  \begin{aligned}
    (\eps_{ee} - \eps_{\mu\mu}) &\to - (\eps_{ee} - \eps_{\mu\mu}) - 2 \,, \\
    (\eps_{\tau\tau} - \eps_{\mu\mu}) &\to -(\eps_{\tau\tau} - \eps_{\mu\mu}) \,, \\
    \eps_{\alpha\beta} &\to - \eps_{\alpha\beta}^* \qquad (\alpha \neq \beta) \,.
  \end{aligned}  
\end{equation}
Eq.~\eqref{eq:osc-deg} shows that this degeneracy implies a change in
the octant of $\theta_{12}$ (as manifest in the LMA-D fit to solar
neutrino data~\cite{Miranda:2004nb}) as well as a change in the
neutrino mass ordering, \textit{i.e.}, the sign of $\Dmq_{31}$. For
that reason it has been called ``generalized mass ordering
degeneracy'' in Ref.~\cite{Coloma:2016gei}.

The $\eps_{\alpha\beta}$ in Eq.~\eqref{eq:NSI-deg} are defined in
Eq.~\eqref{eq:eps-eff} and depend on the density and composition of
the medium. It is easy to see that, if NSI simultaneously affect both
up and down quarks with a coupling proportionally to their charge,
$\eps_{\alpha\beta}^{u,V} = -2\eps_{\alpha\beta}^{d,V}$, the
dependence on $x$ cancels out for neutral matter and the degeneracy is
mathematically exact. In this work, however, we consider only NSI with
either up or down quarks and hence the degeneracy will be approximate,
mostly due to the non-trivial neutron density along the neutrino path
inside the Sun~\cite{Gonzalez-Garcia:2013usa}.

\begin{figure*} \centering
  \includegraphics[width=0.8\textwidth]{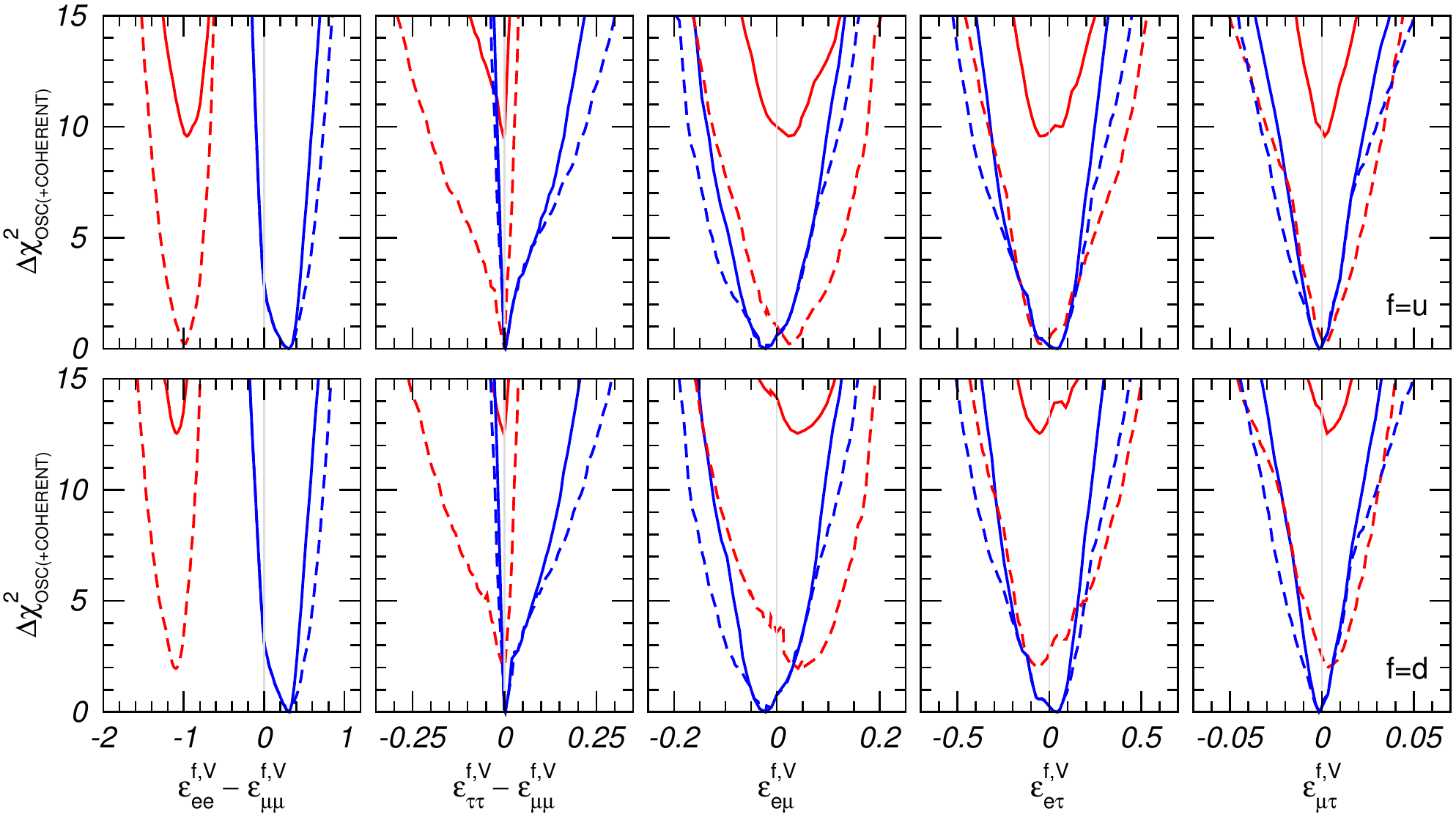}
  \caption{$\Delta\chi^2$ as a function of NSI parameters
    $\epsilon_{\alpha\beta}^{f,V}$, for a global fit to oscillation
    experiments (dashed curves) and for a fit to oscillations and
    COHERENT data (solid curves). Blue lines correspond to the LMA
    solution ($\theta_{12}<\pi/4$), while the red lines correspond to
    the LMA-D solution ($ \theta_{12}> \pi/4$). We minimize the
    $\chi^2$ with respect to all oscillations parameters and all
    un-displayed NSI parameters in each panel.}
  \label{fig:combo}
\end{figure*}

\mysect{Global fit to oscillation data.}
For oscillation constraints on NSI parameters and the detailed
description of methodology and data included we refer to the
comprehensive global fit in the framework of 3-flavour oscillations
plus NSI with up and down quarks performed in
Ref.~\cite{Gonzalez-Garcia:2013usa}.  In principle the analysis
depends on six oscillations parameters plus eight NSI parameters per
$f$ target, of which five are real and three are phases. To keep the
fit manageable in Ref.~\cite{Gonzalez-Garcia:2013usa} only real NSI
were considered and $\Dmq_{21}$ effects were neglected in the analysis
of atmospheric and long-baseline experiments. This renders the analysis
independent of the CP phase in the leptonic mixing matrix. For further
details see Ref.~\cite{Gonzalez-Garcia:2013usa}. For completeness we 
show the results of this fit as dashed lines in
Fig.~\ref{fig:combo}. Two different sets of solutions are shown: dashed 
blue lines corresponding to the LMA solution, and dashed red lines
corresponding to the LMA-D solution, which implies a flipped mass
spectrum and $\theta_{12}$ in the second octant according to
Eq.~\eqref{eq:osc-deg}.

\mysect{COHERENT results.}
The COHERENT collaboration has recently reported the observation of
coherent neutrino--nucleus scattering at
$6.7\sigma$~\cite{Akimov:2017ade}.  The experiment uses neutrinos
produced from pion decay at rest, and a 14.6~kg CsI[Na] detector. Here
we describe our implementation of their constraints on NSI parameters,
following closely Ref.~\cite{Akimov:2017ade}.

At the SNS, the neutrino flux consists of a monochromatic $\nu_\mu$
line coming from $\pi^+ \to \mu^+ \, \nu_\mu$, plus a continuous
spectrum of $\bar\nu_\mu$ and $\nu_e$ from the subsequent $\mu^+$
decay. Hence the total number of coherent scattering events will
receive contributions from the three flux components, $\nu_\mu$,
$\bar\nu_\mu$ and $\nu_e$. We extract the relative contribution
$f_\alpha$ of each flavour to the total number of events from the
shaded histograms of Fig.~S11 in the supplementary material of
Ref.~\cite{Akimov:2017ade} as $f_{\nu_e} = 0.31$, $f_{\nu_\mu} =
0.19$, $f_{\bar\nu_\mu} = 0.50$. For neutrinos of flavor $\alpha$
interacting with a nucleus with total zero spin, for which both the
sum of proton spins and of neutron spins is also zero, the interaction
rate is sensitive to the following combination of SM and NSI vector
couplings (see, \textit{e.g.,} Ref.~\cite{Barranco:2005yy}):
\begin{multline}
  \label{eq:Qw}
  \hspace{-3mm}
  Q_{w\alpha}^2 \propto \left[
    Z (g_p^V+2\eps_{\alpha\alpha}^{u,V} + \eps_{\alpha\alpha}^{d,V})
    + N(g_n^V +\eps_{\alpha\alpha}^{u,V} + 2\eps_{\alpha\alpha}^{d,V})
    \right]^2
  \\
  + \sum_{\beta\neq\alpha} \left[
    Z (2\eps_{\alpha\beta}^{u,V} + \eps_{\alpha\beta}^{d,V})
    + N(\eps_{\alpha\beta}^{u,V}+2\eps_{\alpha\beta}^{d,V})
    \right]^2 \,.
\end{multline}
Here, $N$ and $Z$ are the number of neutrons and protons in the target
nucleus (we take into account the contributions from both, Cs and I), and $g_p^{V} = 1/2 -
2\sin^2\theta_W$, $g^{V}_n = - 1/2$ are the SM couplings of the $Z$
boson to protons and neutrons, respectively, $\theta_W$ being the weak
mixing angle. The predicted number of signal events $N_\text{NSI}$,
for a given set of NSI parameters $\varepsilon$, can be expressed as:
\begin{equation}
  N_\text{NSI} (\varepsilon)
  = \gamma \left[
    f_e Q_{w e}^2 (\varepsilon)
    + (f_{\nu_\mu} + f_{\bar\nu_\mu})Q_{w \mu}^2 (\varepsilon)
    \right] \,,
\end{equation}
where $\gamma$ is an overall normalization constant which depends on
the exposure, detector efficiencies, etc.
We then construct a chi-squared function $\chi^2_\text{COH}$ using
just the total number of events, according to the expression given in
the supplementary material of Ref.~\cite{Akimov:2017ade}. We consider
$N_\text{meas} = 142$ observed events and take into account the
statistical errors of the signal and the subtracted background, as
well as systematic errors of the signal (28\%) and beam-on background
(25\%). The normalization constant $\gamma$ (which is not given in
Ref.~\cite{Akimov:2017ade}) is determined by requiring the $\chi^2$ to
be zero at the best-fit point quoted in Ref.~\cite{Akimov:2017ade}
(\textit{i.e.}, $\epsilon_{ee}^{u,V} = -0.57$, $\epsilon_{ee}^{d,V} =
0.59$, all other $\epsilon^{f,V}_{\alpha\beta} = 0$).\footnote{Let us
  note that, with this procedure, our constraints on
  $\epsilon_{ee}^{u,V}$ and $\epsilon_{ee}^{d,V}$ turn out slightly
  weaker than the result in Ref.~\cite{Akimov:2017ade} (our 90\%~CL
  interval is about 20\% larger). Hence our results can be regarded as
  conservative.}

\begin{figure} \centering
  \includegraphics[width=0.94\linewidth]{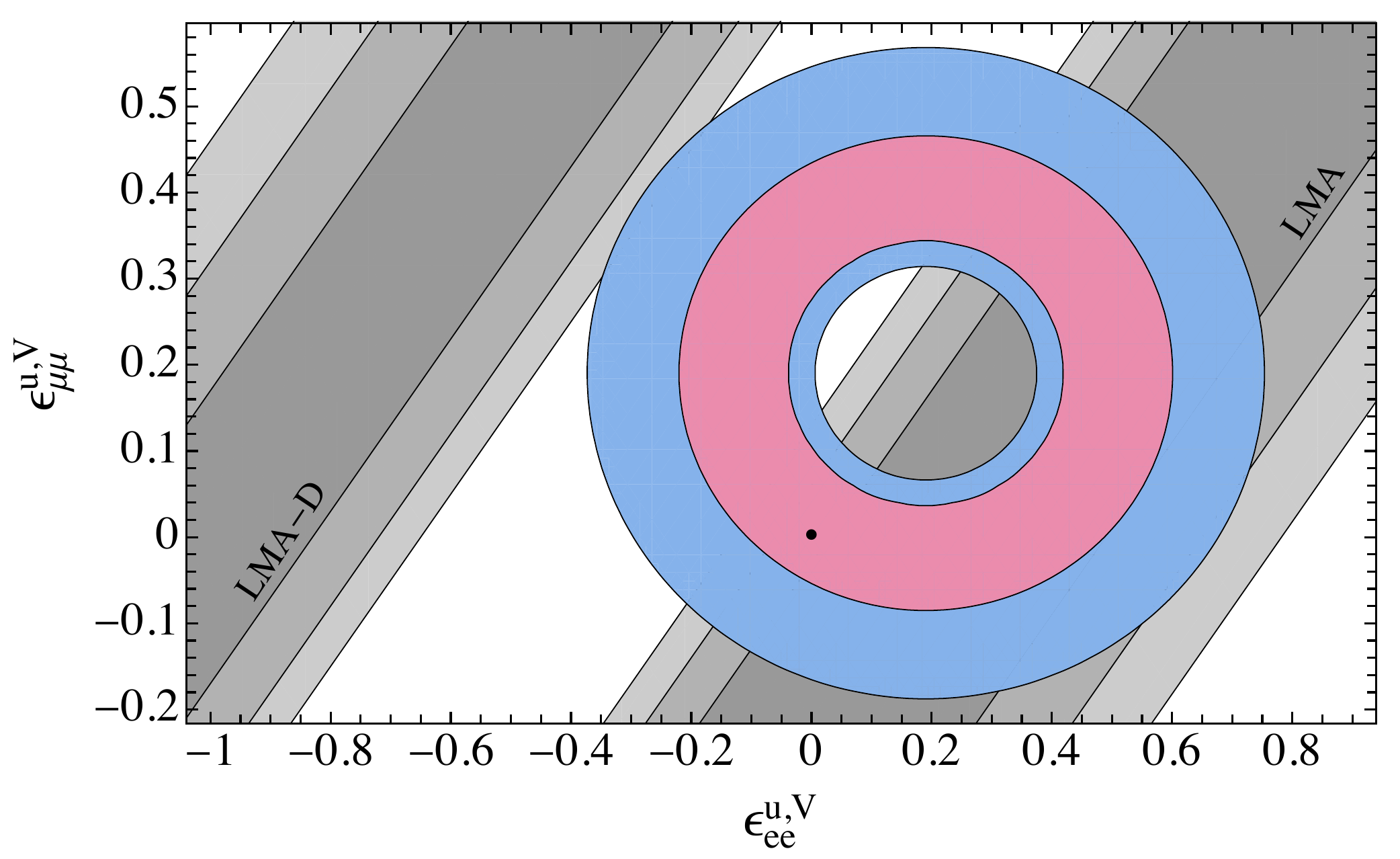}
  \caption{Allowed regions in the plane of $\eps^{u,V}_{ee}$ and
    $\eps^{u,V}_{\mu\mu}$ from the COHERENT experiment shown together
    with the allowed regions from the global oscillation analysis.
    Diagonal shaded bands correspond to the LMA and LMA-D regions as
    indicated, at $1\sigma$, $2\sigma$, $3\sigma$ (2~dof). The
    COHERENT regions are shown at $1\sigma$ and $2\sigma$ only because
    the $3\sigma$ region extends beyond the boundaries of the
    figure.}
  \label{fig:ee-mm}
\end{figure}

To illustrate the impact of COHERENT on the LMA-D solution, we show in
Fig.~\ref{fig:ee-mm} the chi-squared for oscillations and for the
COHERENT experiment separately, projected onto the
$\epsilon_{ee}^{f,V}$ vs $\epsilon_{\mu\mu}^{f,V}$ plane. In this
example, we have restricted to flavour diagonal NSI with $f=u$ quarks.
Oscillation data only constraints the difference $\epsilon_{ee}^{f,V}
- \epsilon_{\mu\mu}^{f,V}$ and therefore two separate bands in this
plane are allowed by the data: one corresponding to the LMA, and a
second one for the LMA-D solution. Conversely, the COHERENT experiment
constrains the combination given in Eq.~\eqref{eq:Qw} and therefore
its results project onto an ellipse in this plane.

\mysect{Results.}
Our final results for the combined fit of oscillations and COHERENT
data are given in Fig.~\ref{fig:combo}, where we show as full lines
the total $\Delta\chi^2 = \Delta(\chi^2_\text{OSC} + \chi^2_\text{COH})$ as a
function of the NSI parameters $\epsilon_{\alpha\beta}^{f,V}$, for
$f=u$ (upper panels) and $f=d$ (lower panels) after marginalization
over the undisplayed oscillation and NSI parameters in each panel.
While the LMA-D solution is perfectly compatible with oscillation data
alone we find that, once COHERENT data is included in the fit, it is
disfavored with respect to LMA with $\Delta\chi^2 \geq 9.6$ (12.6)
for $f=u$ ($f=d$), which corresponds to $3.1\sigma$ ($3.6\sigma$) for
1~dof.

When oscillation parameters are marginalized within the ``standard''
LMA region the global analysis slightly favors non-vanishing diagonal
NSI. The reason for this lies in the $2\sigma$ tension between the
determination of $\Dmq_{21}$ from KamLAND and solar neutrino
experiments (see, for example, Ref.~\cite{Esteban:2016qun} for the latest
status on this issue).

\begin{figure} \centering
  \includegraphics[width=0.94\linewidth]{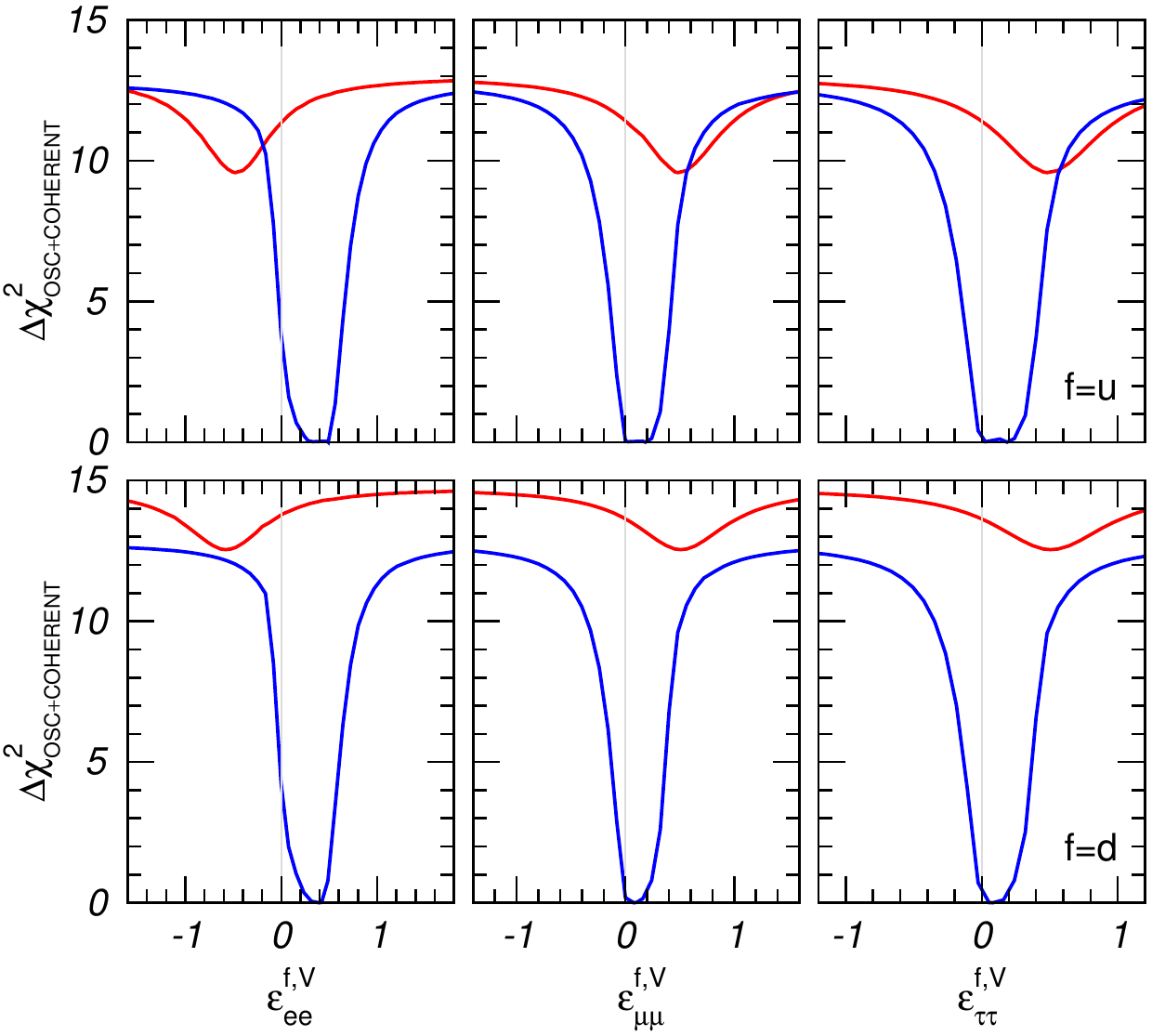}
  \caption{Bounds on the flavour diagonal NSI parameters from the
    global fit to oscillation plus COHERENT data.  Blue lines
    correspond to the LMA solution ($\theta_{12} < \pi/4$), while the
    red lines correspond to the LMA-D solution ($ \theta_{12} >
    \pi/4$).}
    \label{fig:combo2}
\end{figure}

In order to stress the effect of COHERENT in the fit with respect to
the constraints already provided by oscillation data, the results for
the diagonal NSI parameters are shown in Fig.~\ref{fig:combo} for the
differences $\epsilon_{ee}^{f,V} - \epsilon_{\mu\mu}^{f,V}$ and
$\epsilon_{\tau\tau}^{f,V} - \epsilon_{\mu\mu}^{f,V}$, to which
oscillations are sensitive. Notice, however, that the inclusion of
COHERENT data allows to set independent bounds on all
$\epsilon_{\alpha\alpha}^{f,V}$, since COHERENT depends on a different
combination of $\epsilon_{ee}^{f,V}$ and $\epsilon_{\mu\mu}^{f,V}$. We
show the projection of the marginalized $\Delta\chi^2$ for each
flavour diagonal NSI in Fig.~\ref{fig:combo2}. As can be seen, the
combined fit of COHERENT and oscillation data is capable of
constraining the individual flavour diagonal NSI up to 
$\Delta\chi^2\sim 12$ .  Beyond that level oscillation data dominate
and only the two differences relevant for oscillations are effectively
bounded, which leads to the flattening of the marginalized
$\Delta\chi^2$ as a function of the individual diagonal NSI.

\begin{table} \centering
  \begin{tabular}{|c|c|c|}
    \hline\hline
    & $f=u$ & $f=d$ \\ \hline
    $\eps_{ee}^{f,V}$   &  $[0.028, 0.60]$  & $[0.030,0.55]$ \\
    $\eps_{\mu\mu}^{f,V}$&  $[-0.088, 0.37]$ & $[-0.075, 0.33]$ \\
    $\eps_{\tau\tau}^{f,V}$& $[-0.090, 0.38]$ & $[-0.075, 0.33]$ \\ 
    $\eps_{e\mu}^{f,V}$  & $[-0.073,0.044]$  & $[-0.07, 0.04]$ \\
    $\eps_{e\tau}^{f,V}$ & $[-0.15, 0.13]$ & $[-0.13, 0.12]$ \\
    $\eps_{\mu\tau}^{f,V}$ &  $[-0.01, 0.009]$ & $[-0.009, 0.008]$\\
    \hline\hline
  \end{tabular}
  \caption{Allowed ranges at 90\% CL for the NSI parameters
    $\eps_{\alpha\beta}^{f,V}$ for $f=u,d$, as obtained from a global
    fit to oscillation and COHERENT data.  The results for each NSI
    parameter are obtained after marginalizing over all oscillation and
    the other NSI parameters.}
  \label{tab:90CL}
\end{table}

The 90\% CL allowed ranges for the NSI parameters from our global
analysis are given in Tab.~\ref{tab:90CL}. The addition of COHERENT
data allows to derive competitive constraints on each of the diagonal
parameters separately.  This is especially relevant for
$\epsilon^{f,V}_{\tau\tau}$ for which the new bound $-0.09
<\epsilon_{\tau\tau}^{u,V} < 0.38$ ($-0.075 <
\epsilon_{\tau\tau}^{d,V} < 0.33$) at 90\%~CL represents the first
direct bound on NC vector interactions of $\nu_\tau$ assuming light
mediators and is an order of magnitude stronger than previous indirect
(loop induced) limits~\cite{Davidson:2003ha}.  We also see that for
$\epsilon^{f,V}_{ee}$ the 90\% CL range does not include zero. As
explained above this ``non-standard'' result is driven by by the
$2\sigma$ tension in the determination of $\Dmq_{21}$ from KamLAND and
in solar neutrino experiments.

\mysect{Conclusions.}
In this Letter, we have combined the recently reported measurement of
neutrino--nucleus coherent scattering by the COHERENT collaboration
with data from neutrino oscillation experiments, in order to constrain
neutrino NSI affecting NC interactions with quarks. We find that the
addition of COHERENT to the global fit from oscillation data excludes
the LMA-D solution at $3.1\sigma$ ($3.6\sigma$) CL for NSI with up
(down) quarks.  In addition, the combination of oscillation and
COHERENT data allows to derive competitive constraints on all diagonal
NSI parameters individually.

\mysect{Acknowledgements.}
This work is supported by USA-NSF grant PHY-1620628, by EU Networks
FP10 ITN ELUSIVES (H2020-MSCA-ITN-2015-674896) and INVISIBLES-PLUS
(H2020-MSCA-RISE-2015-690575), by MINECO grants FPA2016-76005-C2-1-P,
FPA2012-31880 and MINECO/FEDER-UE grants FPA2015-65929-P and
FPA2016-78645-P, by Maria de Maetzu program grant MDM-2014-0367 of
ICCUB, by the ``Severo Ochoa'' program grant SEV-2016-0597 of IFT.
This manuscript has been authored by Fermi Research Alliance, LLC
under Contract No. DE-AC02-07CH11359 with the U.S. Department of
Energy, Office of Science, Office of High Energy Physics. The
publisher, by accepting the article for publication, acknowledges that
the United States Government retains a non-exclusive, paid-up,
irrevocable, world-wide license to publish or reproduce the published
form of this manuscript, or allow others to do so, for United States
Government purposes.

\bibliographystyle{JHEP}
\bibliography{NSI}

\end{document}